\documentclass[mathleft
]{an}
\usepackage{graphicx}
\usepackage{times}
\begin{document}


\title{Effects of Turbulent Viscosity on A Rotating Gas Ring Around A Black Hole: Results in Numerical Simulation}

\author{Kinsuk Giri\inst{1}\fnmsep\thanks{Corresponding author:
\email{kinsuk@mx.nthu.edu.tw}\newline}
\and  Hsiang-Kuang Chang\inst{1}
}
\titlerunning{Effects of Turbulent Viscosity on A Rotating Gas Ring}
\authorrunning{Kinsuk Giri \& H.K. Chang}
\institute{
Institute of Astronomy, National Tsing Hua University, 101 Section 2, Kuang Fu Road, Hsinchu 30013, Taiwan, ROC}

\publonline{later}


\abstract{
In this paper, we present the time evolution of a rotationally axisymmetric gas ring around a non rotating black hole
using two dimensional grid-based hydrodynamic simulation. We show the way in which angular momentum transport is included
in simulations of non-self-gravitating accretion of matter towards a black hole.
We use the Shakura-Sunyaev $\alpha$ viscosity prescription to estimate the turbulent viscosity  for all major viscous stress tensors. 
We investigate how a gas ring which is initially assumed to rotate with Keplerian angular velocity is accreted
on to a black hole and hence forms accretion disc in the presence of turbulent viscosity.
We show that the centrifugal pressure supported sub-Keplerian flow with shocks forms when the ring starts to disperse
with inclusion of relatively small amount of viscosity. But, if the viscosity is above the critical value, the shock disappears
altogether and the whole disc becomes Kepleiran which is subsonic everywhere except in a region close
to the horizon, where it supersonically enters to the black hole.
We discovered a multiple valued Mach number solution and the corresponding density distributions 
that connects matter (a) from the initial Keplerian gas ring to a sub-Keplerian disc with shocks in presence of small amount of viscosity
and (b) from the sub-Keplerian flow to a Keplerian disc in presence of huge amount of viscosity.
We calculate the temporal variations of the  magnitude of various time scales which ensure us
about the stability of the flow.}
\maketitle

\section{Introduction}
The main source of energy in many astrophysical objects is accretion which includes
different types of binary stars, binary X-ray sources, most probably quasars and active galactic nuclei (AGN).
Intensive development of the accretion disk theory begins after the first discovery of the extra solar X-ray sources
by Giacconi and his team (Giacconi et al. 1967) in which accretion was the only possible way.
There exists a few dominating factors that determine the morphology of an accretion flow.
In general, some matter is assumed to rotate in Keplerian orbits inside an accretion disc.
But it is only possible when it is in equilibrium, i.e., gravity is balanced by
the centrifugal force and gas is sufficiently cold and has only rotational motion. If
this is the case, then nothing would have ever happened inside the accretion disc. In
this situation, matter would just go on revolving around the compact object forever.
However, this is not what happens because of viscosity which transports momentum, and therefore angular momentum.
Viscous processes are of great importance in the theory of accretion discs. In order to move
radially inwards, the disc material has to lose it's angular momentum. Therefore a
steady flow of angular momentum radially outwards and of mass radially inwards takes place in the disc.
But, the main source of this large amount of viscosity is still a mystery in this subject.
The kinematic viscosity $\eta$ has a dimension of $L^2 /T$, where $L$ is length and $T$ is time.
Diffusion at the molecular level generates viscosity, and is called
the molecular viscosity for which the length $L$ is comparatively very small.
Evidently, the molecular viscosity is microscopic in nature and caused by
frictious and dragging interactions of individual neighbouring particles.
So, the molecular viscosity fails to explain the cause of advection of matter towards the compact object.
We therefore need to identify possible instabilities that can cause a turbulence which generate huge viscosity
larger than that caused by the molecular viscosity.
Hence, it is clear that some kind of macroscopic turbulent viscosity
must be present. A significant factor is how fast the angular momentum of accreting
matter can be eliminated. However, a very successful idea of parametrizing the viscosity without
identifying its source was given in 1973 by Shakura \& Sunyaev (hereafter SS73) who first proposed the so-called $\alpha$-
parameter to measure the turbulent viscosity which is used to determine the efficiency of angular momentum transport.
Later, analytically, it has been shown that the turbulent viscosity is an important parameter that controls the
ability of the accreting disc to produce  the inward
transport of mass and outward transport of angular momentum (Lynden-Bell \& Pringle, 1974; Pringle, 1981).

A few works have been done on the time evolution of a rotating viscous gas ring
which is used as a testing model for numerical methods developed to simulate accretion discs (Speith \& Riffert 1999, Kley 1999).
However, these works largely discuss the properties of the transonic solutions which do not includes shocks that
is generated in the flow due to fighting between  gravity and centrifugal forces (Chakrabarti 1989).
In the presence of sufficient angular momentum, slowly moving matter at the outer edge of the ring/disc gradually
gains radial velocity while getting accreted towards the compact object and becomes
super-sonic. At some point, super-sonic matter may make discontinuous transition
to the sub-sonic branch. This discontinuous transition is known as a shock.
Early known simulations were also carried out by Hawley \& Smarr (1986) shows the transonic solutions with shocks.
The dynamics of a rotationally symmetric viscous gas ring around a Kerr black hole is calculated in
the thin-disk approximation (Riffert 2000). However, because their main objective was much wider, they
did not elaborate on the estimations and effects of turbulent viscosity (SS73) on a rotating ring.
It is noted that MHD turbulence considered in  magneto-rotational instability(MRI) is obviously a
phenomenon that we cannot hope to capture in a hydro simulation, however we can simply introduce
a shear viscosity term (SS73) into our numerical scheme to simulate its effect in angular momentum transportation.

The path breaking model in SS73 which is commonly known as 'Keplerian disc' successfully
explains the multi-colour black body component spectrum which is observed in the
spectra of several accretion disk in the galactic and extra-galactic system but unable
to produce power low emission at the higher energy limit shown by most of the black
hole candidates. More importantly, in this model (SS73), accretion disc is terminated at
the marginally stable orbit and the inner boundary conditions are not satisfied since
the advection of flow was completely ignored. It is also fact that the basic requirement of
the accretion process around black hole, i.e., the transonic property, is not satisfied
in this model. Therefore the study of the properties of the sub-Keplerian 
component which include shocks or no shocks becomes as essential, and perhaps
more essential than the Keplerian component itself.
On the theoretical side, since the pioneering work by SS73 thin disc models, parametrized
by the so-called turbulent or $\alpha$ viscosity has been applied successfully to many numerical works in accretion disc solutions, viz.
Robertson \& Frank 1986, Chakrabarti 1989, 1990, Chakrabarti \& Molteni 1995, Igumenshchev et al. 2000, McKinney \& Narayan 2007,
Yuan et al. 2012, Giri \& Chakrabarti 2012, hereafter GC12, Kumar \& Chattopadhyay 2013, Giri \& Chakrabarti  2013,
Giri et al. 2015.
In GC12, we study the $\alpha$ viscosity effects on  inflowing
sub-Keplerian matter towards a black hole. In that paper we mainly concentrated on the time evolution of matter which
was initially assumed to rotate with sub-Keplerian velocity at the outer boundary of our simulation.
But, till now, however, we did not study on the time evolution of a gaseous ring which is initially
assumed to rotate around a black hole with Keplerian velocity.
In the present paper, we have shown how the turbulent viscosity plays a key
role in advection of the matter towards a black hole from an initial gas ring. We also, therefore, find the existence and disappearance
of the shock  solutions near black hole which is controlled by the magnitude of turbulent viscosity.
The plan of this paper is in next paragraph.

In the next section, we show how we introduce the governing equations that we have solved by a
the grid-based finite difference method. In sections 3 and 4, we present the methodology
and the results of our simulation, respectively. In next section, we discuss on time scales and stability.
Finally, in section 6, we draw concluding remarks.

\section{Governing Equations}
A continuum physical system is described by the laws of conservation of mass, momentum and energy.
The conservation of mass of the flow is described by the continuity equation
for the density $\rho$ and flow velocity ${\bf v}$ which is given by (Landau \& Lifshitz 1959)
\begin{equation}
{\partial \rho \over \partial t}+ {\bf \bigtriangledown} . (\rho {\bf v}) = 0.
\end{equation}
The momentum conservation are given by the {\it Navier-Stokes equation}:
\begin{equation}
{\partial {\bf v} \over \partial t} + {\bf v}.{ \bf \bigtriangledown} {\bf v}
 = - {1 \over \rho}{\bf \bigtriangledown} P + {\bf f_{external}},  
\end{equation}
where, $P$ is the gas pressure at each point arising because of the thermal motion
of the gas particles and ${\bf f_{external}}$ denotes the external forces like gravity, viscosity, body
forces etc. The energy equation for the gas element is given by,
\begin{equation}
{\partial \over \partial t}({ 1 \over 2}\rho v^2+\rho {\cal E}) +
{\bf \bigtriangledown} . [({1 \over 2} \rho v^2 + \rho {\cal E} + P) {\bf v}]= 
{\bf f.v} +\Omega-\Gamma
\end{equation}
where, the terms $\rho v^2$ and $\rho {\cal E}$ measure the kinetic energy density and internal energy
density respectively. $\Omega$  and $\Gamma$ are cooling and heating term respectively.
In our work we have dealt with adiabatic astrophysical flows without explicit consideration of radiative transfer
effects.  So, in our present work, we have neglected the cooling and heating terms.

The mass, momentum and energy conservation equations to describe a 2D axisymmetric inviscid flow around a
Schwarzschild black hole in a compact form using non-dimensional 
units are already presented in Giri et al. (2010, hereafter G10). The self-gravity of the accreting
matter is ignored. Cylindrical coordinate $(x, \phi, z)$ is adopted
with the $z$ axis being the rotation axis of the disc. 
Instead of using general relativity, we use the well known pseudo-Newtonian
potential prescribed by  Paczy\'{n}ski \& Wiita (1980, hereafter PW80).
We use the mass of the black hole $M_{bh}$, the velocity of light $c$
and the Schwarzschild radius $r_g = 2GM/c^2$
as the units of the mass, velocity and distance respectively.
From Eqn. 2, in an inertial frame of reference, the modified general form of the equations
of the viscous flow (Batchelor 1967) is
\begin{equation}
\rho [{\partial {\bf v} \over \partial t} + {\bf v} . {\nabla {\bf v}}] = 
- { \nabla P} + {\bf {F_b}}  + {\nabla . {\bf {\tau}}},
\end{equation}
where, ${\bf v}$ is the flow velocity, $\rho$ is the fluid density, $P$ is the pressure,  ${\bf \tau}$ is the stress tensor, and ${\bf {F_b}}$
represents body forces (per unit volume) acting on the fluid and ${\nabla}$ is the Del operator.
Typically body forces consist of only gravity forces, but may include other types.
Here ${\bf {\tau}}$ is the viscous stress having six mutually
independent components. In cylindrical coordinates   the components of the velocity vector given by ${\bf v} = ({v_x}, {v_{\phi}}, {v_z})$.
The six independent components  of the viscous stress tensor (LL59)
are listed here in cylindrical coordinates, ${\tau}_{xx} , {\tau}_{x{\phi}} , {\tau}_{xz},
{\tau}_{{\phi}{\phi}}, {\tau}_{{\phi}{z}}$ \& ${\tau}_{zz}$. 
In the case of a thin accretion flow, it is customary to use only 
${\tau}_{x{\phi}}$  component since it is the dominant contributor to the viscous
stress (SS73). This is responsible for transporting angular momentum along the radial direction.
The other components are assumed negligible. But, it is noted that 
in the case of thick accretion flow, all viscous stress could be significant in flow dynamics (GC12). 
In the present work in order to obtain the viscous effects along all directions, we have taken all the viscous
stress  components.
The dimensionless equations governing the viscous flow in our system have been presented 
in GC12 in great detail and we do not repeat them here.
The $SS\alpha$-model (SS73) introduced a phenomenological shear stress into the equations of motion to model the effects of this
turbulence. This shear stress is proportional to $p$, the total pressure. This shear
stress permits an exchange of angular momentum between neighbouring layers.
Using $SS\alpha$ prescription, if the disc is thin, we can assume
\begin{equation}
 {\tau}_{x\phi}= -{\alpha p}, 
\end{equation}
where, $\alpha$ is a proportionality factor which is assumed not be constant throughout the flow.
 It is true that SS viscosity presciptions was applied
for a thin disk, but if one can also use it for thick flows if the $r\phi$ component is the 
only dominant component.
The 'negative' sign in Eqn. 5 is taken in order to assume the negative shear stress. This is assumed
because negative shear stress helps angular momentum to transport outwards and hence it is necessary for accretion.
In this work, we have used this method to quantify turbulent viscosity.

\section{Methodology and Simulation Setup}
The setup of our simulation is the same as presented in G10 and GC12.
Hence, we are not repeating those once again. 
To model a gaseous ring problem, we consider an axisymmetric rotating ring of flow of gas in the pseudo-Newtonian
gravitational field of a black hole of mass $M_{bh}$ located at the centre in the cylindrical coordinates
$[x,\theta,z]$. We assume that at infinity, the gas pressure is negligible and the
energy per unit mass vanishes. To mimic the general relativistic effects near the black hole,we assume the gravitational field
which is  described by PW80,
\begin{equation}
\phi(r) = -{GM_{bh}\over(r-r_g)}, 
\end{equation}
where, $r=\sqrt{x^2+z^2}$, and the Schwarzschild radius is given by,
\begin{equation}
r_g = {{2GM_{bh}} \over {c^2}}.
\end{equation}
We also assume a polytropic equation of state for the
accreting (or, outflowing) matter, $P=K \rho^{\gamma}$, where,
$P$ and $\rho$ are the isotropic pressure and the matter density
respectively, $\gamma$ is the adiabatic index (assumed
to be constant throughout the flow, and is related to the
polytropic index $n$ by $\gamma = 1 + 1/n$) and $K$ is related
to the specific entropy of the flow $s$.

Our computational box occupies one quadrant of the $x-z$ plane with $0 \leq x \leq 200$ and $0 \leq z \leq 200$.
We have considered a circular ring (an anulas in 2D) at equatorial plane,i.e. $x-y$ plane. The centre of the ring is taken at $r_c=100 r_g$.
Without any loss of generality, the radius of the ring is taken as $20 r_g$.  We have chosen the initial density
of the gas inside the ring is ${\rho}_{ring} = 1$ for convenience since, in the absence of self-gravity
and cooling, the density is scaled out, rendering the simulation results valid for any amount of matter.
In the beginning of the simulation, we considered that the matter inside the ring is rotating with it's Keplerian velocity (${\Omega}_k$) is given by
\begin{equation}
{{\Omega}_k} = [{ 1 \over r}{\partial {\Phi} \over \partial r}]^{1 \over 2}, 
\end{equation}
where, $\Phi$ is the gravitational potential.
Initially, the radial velocity of the gas particle inside
the ring is assumed to be zero, i.e. we assume the ring is rotating
it's Kepelrian orbit. 
We need the  sound speed $a$ (i.e., temperature) of the matter inside the ring.
We assumed the sound speed $a$ is low. In order to mimic the horizon of the black hole at the Schwarzschild radius,
we placed an absorbing inner boundary
at $r = 1.01 r_g$, inside which all material is completely absorbed into the black hole. For the background matter
(required to avoid division by zero) we used a stationary gas with density ${\rho}_{bg} = 10^{-6}$ and sound speed (or temperature) the same as that of the gas ring. All the calculations were performed with $512 \times 512$ cells, so each grid has a size of approximately $0.40$ in units of the Schwarzschild radius.
All the simulations are carried out assuming a stellar mass black hole $(M = 10{M_\odot})$.
The procedures remain equally valid for massive/super-massive black holes.
We carry out the simulations till several thousands of dynamical time-scales are passed.
In reality, this corresponds to a few seconds in physical units.

We numerically solve the set of Hydrodynamic equations shown in above using
a finite difference method based code which uses the principle of Total Variation Diminishing (TVD) which was originally
developed by Harten (1983). TVD is a completely Eulerian numerical scheme, which uses an ensemble of grids
to model a fluid. In the astrophysical context, Ryu et al. (1995) developed the TVD scheme to study astrophysical inviscid flows
around black holes. We incorporate the viscosity in this TVD code.

\section{Numerical Results}
In this section we present the results of the simulations with various input viscosity parameters. 
First, we consider for non-viscous case. Afterwards, the cases for various viscous parameters are investigated.
As mentioned earlier, we chose the outer boundary of the simulation
grid at $r = 200$. The specific angular momentum ($\lambda$) of the initial gas ring is chosen to be Keplerian.
We stop the simulation at $t = 100$ s. This
is more than several thousand times the dynamical time. Thus, the solution has
most certainly come out of the transient regime and started exhibiting solution characteristics of its flow parameters.
The simulation results will be discussed now.
\begin{figure}
\begin{center}
\includegraphics[height= 8.0truecm, width=8.0truecm, angle=0]{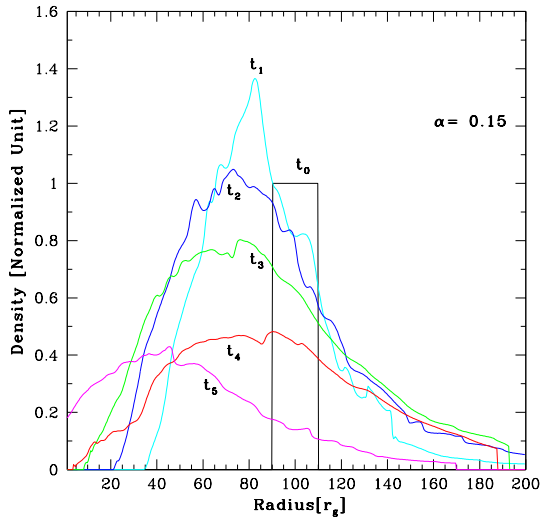}
\caption{Time variation of radial density distribution at six different times: $t = 0 (t_0), 12.5  (t_1), 25 (t_2), 50 (t_3), 75 (t_4), 100 (t_5) s$.
The $\alpha$ parameter is $\alpha = 0.15$.}
\end{center}
\end{figure}

In order to check the effects of turbulent viscosity, for an optimal case, we run our simulation with sufficiently
large amount of $\alpha = 0.15$ .
The behaviour of the time-dependent solution is evident in
To show how the matter gradually advect towards the black hole from a initial gas ring in presence of viscosity,
in Fig.1,  we plot the the density distributions along radial direction at equatorial plane
at regular intervals at times t = 0.0, 12.5, 25, 50, 75 and 100 secs.
It is evident that, the matter from the initial ring gradually advected towards the black hole as viscosity helps to transport the
angular momentum of the flow outwards. The density distributions in the post centre region of the initial ring resemble those of 
analytical solutions of Lynden-Bell \& Pringle (1974), though our results are
more realistic since all of the viscous stress tensors plays important role in our time dependent numerical simulation. 
\begin{figure}
\begin{center}
\includegraphics[height= 8.0truecm, width=8.0truecm, angle=0]{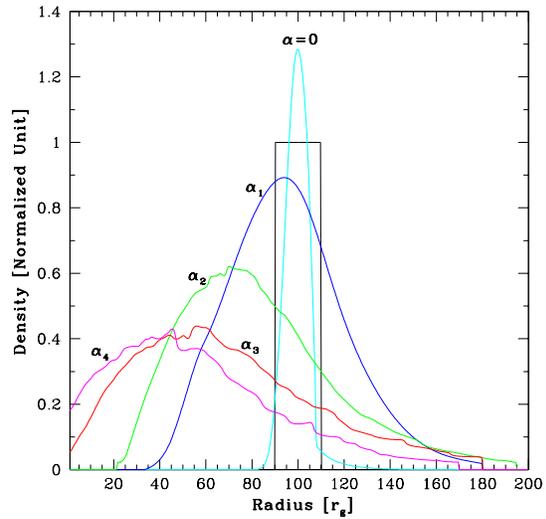}
\caption{Variations of the radial density distribution(in normalized units)
at equatorial plane with the initial distribution (shown in rectangular shape)
centered at $r = 100 rg$ with width $20 rg$. All plots are taken at $t = 100$s. As the viscosity parameter
is increased, the angular momentum is transported outwards and hence matter is advected inwards.
Density profiles are drawn for $\alpha =  0, 0.01, 0.05, 0.1 \& 0.15$.}
\end{center}
\end{figure} 
Next, we run several cases for various viscous parameter $\alpha$ including a run where viscosity in our system is assumed to be zero, i.e $\alpha = 0$.
In order to  distinguish between numerical and viscous effects, we selected various values of $\alpha$ parameters for the simulations,
and we inspected different types of result for a wide range of viscous parameter.
In Fig.2, we compare the density distribution at the equatorial plane of the flow for
various viscous parameter $\alpha$. Each distribution is the time evolution for an initial
distribution centered around $r = 100 r_g $ with width $20 r_g$. To make the comparison meaningful, all the runs
were carried out up to $t = 100 s$. Each result is obtained starting
with an same initial gas ring with Keplerian rotation as mentioned earlier.
The values of $\alpha$ for which the curves are drawn are shown in the figure;
$\alpha = 0$, ${\alpha}_1 = 0.01$,
${\alpha}_2 = 0.05$, ${\alpha}_3 = 0.1$ and ${\alpha}_4 = 0.15$.
\begin{figure}
\begin{center}
\includegraphics[height= 8.0truecm, width=8.0truecm, angle=0]{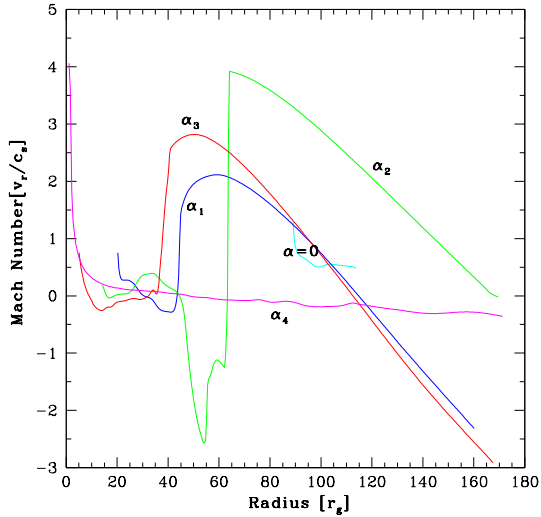}
\caption {Comparison of the radial distribution of Mach numbers obtained from the various
cases for different viscous parameters (marked). The mach number variation on the equatorial plane is shown.
The initial ring reduce to a shocked sub-Keplerian flow for $\alpha = 0.01, 0.05, 0.1$. The shock disappears from the flow  when $\alpha = 0.15$.
It is noted that $\alpha = 0.15$, the entire flow become subsonic except near to black hole horizon where the
flow supersonically enters towards black hole}
\end{center} 
\end{figure}

As the viscosity parameter is increased, the flow behaviour changes dramatically.
The matter from the initial gas ring gradually accretes towards the black hole with enhancement of viscous parameter.
It is noted that when no viscosity is there, even with time evolution,
the matter of the initial gas ring just go on revolving at approximately  same orbit,
although numerical viscosity which is negligible with respect to turbulent viscosity 
slightly changes the shape of the ring.
It is noted that any grid based finite difference method causes this kind numerical viscosity
which is unavoidable. As turbulent viscosity is enhanced, the angular momentum is transported outwards and
this causes in most of the matter moving inwards and eventually accreting towards the black
hole. It is also evident that angular momentum carried outwards by a small amount of material.
\begin{figure}
\begin{center}
\includegraphics[height= 10.0truecm, width=8.0truecm, angle=0]{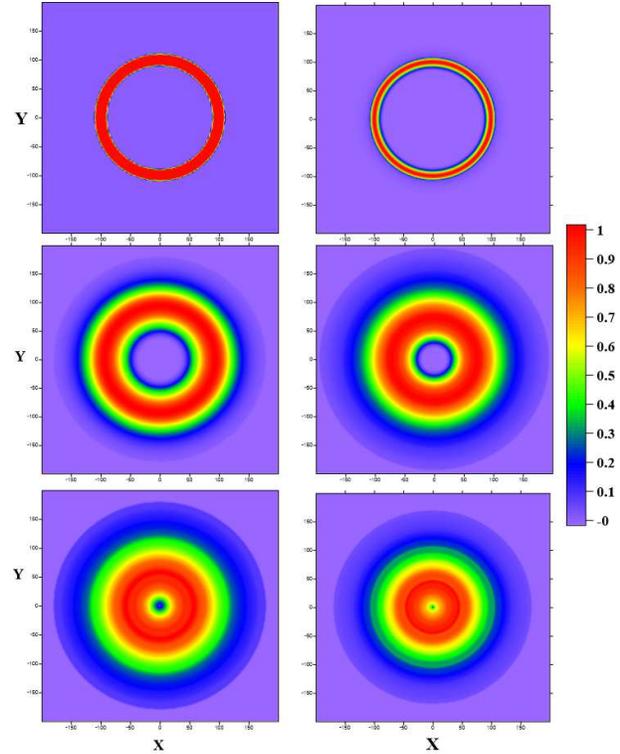} 
\caption{Changes in the density distribution at $x-y$ plane with the change of the viscous parameter $\alpha$
at $t = 100 s$. Here, densities are in normalized unit, 
and radius and velocity are in Schwarzschild unit. All cases are started for the same initial condition (top-left).
The $\alpha$ parameters are $0$ (top-right), $.01$ (middle-left), $0.05$ (middle-right), $0.1$ (bottom-left) \& $0.15$ (bottom-right).
The initial density distribution at $t = 0$ is shown in the top left plot.}
\end{center}
\end{figure}

We now turned our focus on flow dynamics,i.e. the velocity and temperature distribution of the
simulated flow in order to study the sonic property of the flow near the black hole.
In Fig. 3, we compare the Mach number  variations in the equatorial plane of the
flow for various $\alpha$. The Mach number is defined as the ratio of the flow radial velocity to
the sound speed at the same location.
All the runs were carried out up to t = 100 s. Each result is obtained starting
with the same initial gas ring and then gradually increasing
$\alpha$ till the flow supersonically reach up to black hole horizon. 
The values of $\alpha$ for which the curves are drawn in
Fig. 3 are 0.0,0.01,0.05,0.1 and 0.15 for the curves coloured with indigo, blue, green, red and magenta respectively.
For, $\alpha = 0$, the Mach number distribution does not show any kind of specialty since, the ring remains Keplerian.
But, from $\alpha = 0.01$ onwards the Mach number dramatically changes, the flow first become supersonic and suddenly 
jumps to subsonic state, so that shocks form. The strength of shocks become highest when $\alpha = 0.05$ and then it
starts to be weaker and when $\alpha = 0.15$, the shock completely disappears from the system. 
A shock formation necessarily requires that the flow jumps from supersonic to subsonic. 
In this run, the shock disappears for  $\alpha = 0.15$,
which is expected to  the very near to the critical value of $\alpha$ here.
Long ago, Chakrabarti (1990) predicted this type of critical $\alpha$ for a complete sub-Keplerian flow around black holes.
In contrast, in our simulation, we see that even if for an initial Keplerian gas ring, in presence of
relatively small amount of viscosity the ring starts to reduce a disc which behaves like
a sub-Keplerian flow with shock, but, the shocks disappears in presence of excess viscosity which may be greater or equal to the
the ``critical value'' of $\alpha$.
It is also noted that for $\alpha = 0.15$ which contributes a huge amount of viscosity, the whole disc become Keplerian. 
As the Keplerian flow itself is a subsonic flow, the time evolution of the Keplerian flow itself do not produce a shock. 
Hence, the  Keplerian flow remained Keplerian until very close to the black hole where the flows become supersonic and enter to the black hole.
In Fig.3 for each case either shocks or no shocks, we noticed that the solution is transonic which is the necessary property of black hole accretion.
\begin{figure}
\begin{center}
\includegraphics[height= 10.0truecm, width=8.0truecm, angle=0]{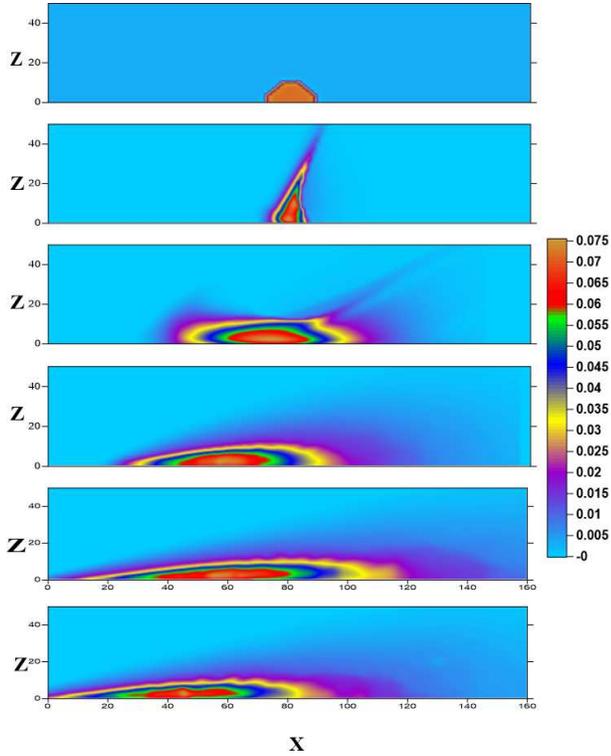}
\caption{Changes in the density distribution at $xz$ plane with the change of the viscous parameter $\alpha$ at $t = 100 s$
. Here, densities are in normalized unit,
and radius and velocity are in Schwarzschild unit. All cases are started for the same initial condition as in (a).
The $\alpha$ parameters are  (b) $0$ , (c) $.01$ (d) $0.05$ (e) $0.1$ \&  (f) $0.15$.}
\end{center}
\end{figure}

In order to make a better representation and visualization, we now draw the two dimensional density maps at Fig. 4.
In Fig.4, we show how the density of matter vary with viscosity through out the $x-y$ plane. The parameters are same
as in Fig.2.
Though we performed our simulation in $x-z$ plane, but, with the help of axisymmetric property  which is assumed in our simulation,
we present the results in $x-y$ plane.
All cases are started for the same initial condition which is shown in(top-left) of Fig.4.
It is clear that for a large amount of viscosity, the density maximum of the gas ring moves toward the black hole, and
at the same time some mass is moving outward, and subsequently the initial gas ring spreads toward larger radii.
For the case where $\alpha = 0$, the gas ring (top right in Fig. 4) remain almost unaltered even at the end of the simulation.
A very slight change in shape is caused due to the artificial numerical viscosity which is negligible related
to the turbulent viscosity  present in the numerical code. 
Though, this artificial viscosity is negligible in contrast of turbulent viscosity.

It is interesting to see the density distributions in vertical directions, i.e. throughout the  $x-z$ plane for the previous cases.
As we initially assume the very thin ring and we have taken $\alpha$ prescription for thin disc,
it is expected that the simulated disc will be thin.
In Figs. 5(a)-(f), we show the distributions of the density
of the flow at the end of our simulation for different values of $\alpha$.
For comparison, we keep the the initial distribution (i.e. at $t = 0$) at Fig. 5(a).
Each of the results of Fig. 5(b-f) are also shown at $t = 100 s$. The values of $\alpha$ are
$0.0$, $0.01$, $0.05$, $0.1$ and $0.15$ for Fig. (b), (c), (d), (e) and (f) respectively.
This figure shows that the presence of large amount of viscosity is essential to advect the matter inwards.
\begin{figure}
\begin{center}
\includegraphics[height= 7.0truecm, width=8.0truecm, angle=0]{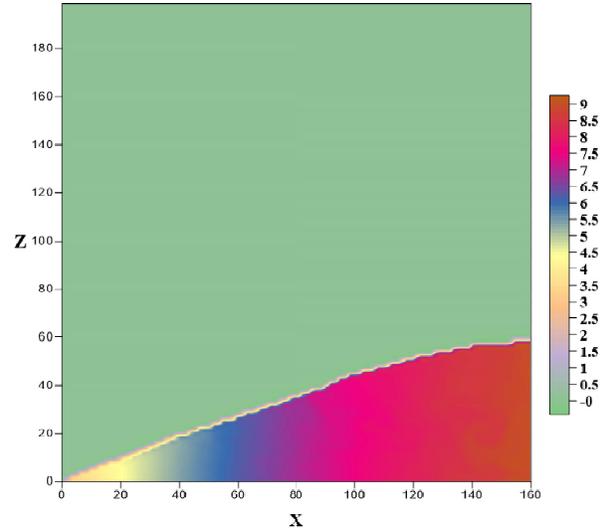}
\caption{Specific angular momentum distribution at $x-z$ plane in our simulation. The parameters are same as in Fig. 5(f).
It is evident that the rotation profile is roughly constant along vertical direction for each $x$.}
\end{center}
\end{figure}
In order to focus on the rotation profile along vertical direction in our simulated disc for large $\alpha$, in Fig. 6,
we plotted the specific angular momentum distribution ($\lambda = v_{\phi} x$) in $x-z$ plane. The parameters of Fig. 6 is
same as in Fig. 5(f). The dimensionless color scale is at the right hand side of the figure.
From Fig. 6, it is clear that for each $x$ the rotational  velocity along vertical direction is constant
which is the characteristics of thin disc i.e. so called `Keplerian disc'.

\section{Timescales and Stability}
Our time-dependent  simulation  is constructed by  $\alpha$ parameter viscosity 
prescription for thin disc (SS73) and its flow pattern is controlled by the amplitude of the turbulent viscosity.
So,the  observations of time-dependent disc behaviour offer one of the few sources of
quantitative information, it is important to calculate the relative magnitudes of the various timescales over which
accretion discs form from an initial gas ring and the effect of eddy viscosity observed on timescales.
Hence, we calculate the three type of typical timescales 
using our simulated results.
In our system, the dynamical timescale ($t_{dyn}$), the timescale on which inhomogeneities on the disk surface rotate is
given by
\begin{equation}
t_{dyn} = \int_{r_{out}}^{r_{in}}  {dr \over v_{\phi}},
\end{equation} 
where, $r_{out}$ and $r_{in}$ are the outer and inner boundary of our simulation box respectively; while 
$v_{\phi}$ is the vertically averaged angular velocity along our simulated thin disc.  
The role of viscosity can be ascertained if one compares
the viscous time with the dynamical timescale. The viscous
timescale ($t_{vis}$), the timescale on which matter diffuses
through the disk under the effect of viscous torques, can be defined as
\begin{equation}
t_{vis} = \int_{r_{out}}^{r_{in}} {{\rho r dr} \over \mu },
\end{equation}
where, $\rho$ is vertically averaged mass density and $\mu$ is the dynamical viscosity coefficient is described by
\begin{equation}
\mu = \alpha \rho {a^2 \over {\Omega}}, 
\end{equation}
where, $\alpha$ is the viscous parameter, $a$ is the adiabatic sound speed, and
\begin{equation}
\Omega = [{ 1 \over r}{{\partial {\Phi}} \over \partial r}]^{1 \over 2}
\end{equation}
is the Keplerian angular velocity. Since our simulation gives roughly Keplerian angular velocity
distribution, we used our simulated angular velocities instead of $\Omega$. 
It is noted that in eqn. 11, we had to supply only viscosity coeffcient $\mu$. 
In order to approximate $\mu$, we used this well known prescription althogh this formula is no
longer valid for non-Keplerian flow.
In the other hand, thermal timescale, the timescale for re-adjustment to thermal
equilibrium, is given by 
\begin{equation}
t_{thermal} \sim {h^2 \over r^2} t_{vis}.
\end{equation}

\begin{figure}
\begin{center}
\includegraphics[height= 8.0truecm, width=8.0truecm, angle=0]{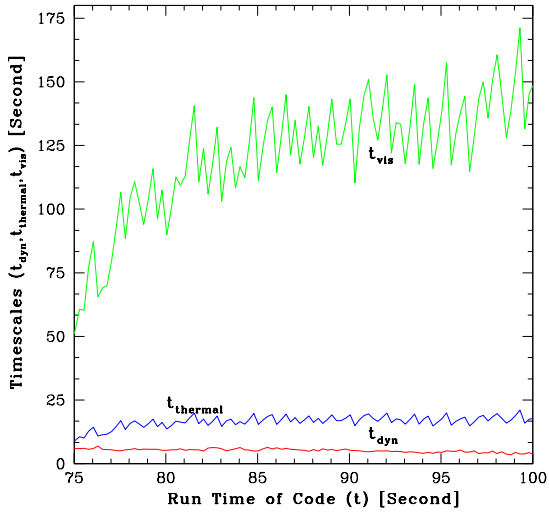}
\caption{Time evolution of the magnitude of dynamical, thermal and viscous (red, blue and green respectively) time scales.
Here, $\alpha = 0.15$ for all the plots.}
\end{center} 
\end{figure}
In Fig. 7, We show the time evolution of the  magnitudes of all the above said time scales in the last quarter of time phase
($75s$ to $100s$)of the run of our simulation. The red curve is for the dynamical time scales, while, blue and green curves
represent thermal and viscous time scales respectively. 
 All the time scales are for the highest viscosity case, i.e. $\alpha = 0.15$.
It is noted that,  we have chosen the last phase of time interval ($75-100$) since in this time spam
our simulation roughly saturates and it behaves like in a  steady state. In Fig. 7, it is clear that 
\begin{equation}
{\bf t_{dyn} < t_{thermal} \ll t_{vis}}.
\end{equation}
Since our disc is thin, this  condition $14$ shows that the centrifugal balance in the radial direction
and hydrostatic balance in the vertical  direction are very rapidly achieved,
while the disc temperature generally evolves on a very longer timescale.

\section{Summary}
Our investigation of the time evolution of the flow dynamics of a rotating gas ring (Keplerian rotation) around a black hole
in the presence of viscosity is based on the different values of $\alpha$ of turbulent viscosity.
We have shown that, in the absence of viscosity, an initially rotating gas ring revolves in same orbit around a black hole
without being advected towards the black hole. We find, however, that in all viscous cases where turbulent viscosity is present,
the rotating ring gradually advected towards the black hole and forms a circular disc.
The results of our simulations show the good agreement with the standard
viscous theory in accretion. We described that the simulations completed with zero viscosity or very low
viscosity gave results that were qualitatively different from more viscous calculations.
We find that the matter from the initial gas ring move inwards as the viscosity is enhanced and the
whole region roughly attains a Keplerian distribution. When the value of viscosity parameter is reasonably large,
the accreting matter reaches up to marginally stable orbit ($\sim 3 r_g$) which is close to the black hole
and the whole disc becomes a Keplerian disc. So, the action of viscosity on the initial gas ring helps most of
the mass moves inwards losing energy and angular momentum and a few percentage of matter  moves outward
to larger radii in order to take up the transported angular momentum.
The radial profiles of density obtained by numerical simulations are in good agreement with standard results.
We also found that the condition of standing gas ring, i.e. the matters in same circular orbit,
satisfied only in a narrow range of the viscous parameter ($0$ to $\sim 0.01$).

It is natural that when angular momentum is too small then the centrifugal force is not
enough to form the shock and the flow is essentially a Bondi type flow (Bondi, 1952).
In our work, from the Mach number variation along radial direction, we found that even when less amount of viscosity is included,
the transported angular momentum need not be sufficient to produce very
strong shock. That is why, $\alpha = 0.01$ did not produce a strong shock.
However, while for $\alpha = 0.05$, the shock is very strong and remain standing there. But, the shock again starts to
become weaker when $\alpha = 0.1$. Finally, the shock propagated backward and disappears for $\alpha = 0.15$ such that
the whole disk become again Keplerian. The  the shock  propagates to large distance making
the whole post-shock region a Keplerian disc. Only region between the horizon and the inner sonic point ($\sim 2 r_g$) of the black hole
will remain supersonic and sub-Keplerian. In general, it is observed that for
a purely sub-Keplerian flows with inclusion of viscosity, the shocks are
weaker (Chakrabarti \& Molteni, 1995), form farther away and are wider as compared to the shocks in inviscid flows.
In this work,  we show that the shock first starts to become stronger with even with 
the increment of $\alpha$, but, after a reasonably large value of
$\alpha$, the shocks starts to become weaker and ultimately disappear when $\alpha$ reaches
to it's critical value. It is expected that in reality
the magnitude of viscosity parameter would be dynamic in nature for a accretion flow dynamics.
So, the variation of shock's nature due to variations of the magnitude viscosity may play an
important role to determine the flow geometry which is essential
to study in temporal and spectral properties of black hole candidates.
It is evident from our simulation that the momentum conservation is so important in accretion disc formation that may run
for thousands of dynamical time scales. We calculated the magnitude of  different time scales in our simulated results to
show the stability of the simulated thin disc.
It is fact that the stability analysis can be done more accurately 
only in three-dimensional simulation which can be extended from our present two-dimensional work. However, this is outside the scope
of the present paper, and will be investigated in future.

\section{Acknowledgments}
This work was supported by the Ministry of Science and Technology, Taiwan, ROC
through grants NSC 101-2923-M-007 -001 -MY3, and 102-2112-M-007 -023 -MY3.

\end{document}